\documentclass[11pt]{article}
\usepackage{amssymb}
\newtheorem{Lemma}{Lemma}
\newtheorem{Theorem}[Lemma]{Theorem}

\newtheorem{Corollary}[Lemma]{Corollary}
\newtheorem{Remark}[Lemma]{Remark}
\newtheorem{Definition}[Lemma]{Definition}

\newenvironment{Proof}
{\begin{trivlist} \item[]{\bf Proof. }}%
{\hspace*{\fill}$\rule{.3\baselineskip}{.35\baselineskip}$\end{trivlist}}

\newcommand{\C}{\mathbb{C}}

\newcommand{\R}{\mathbb{R}}
\newcommand{\Z}{\mathbb{Z}}

\newfam\bifam
\font\tenbi=cmmib10 scaled \magstep1
\font\sevenbi=cmmib10 at 11pt
\font\fivebi=cmmib10 at 6pt
\textfont\bifam = \tenbi
\scriptfont\bifam= \sevenbi
\scriptscriptfont\bifam= \fivebi

\usepackage[dvips]{epsfig}
\usepackage{graphicx}

\begin{document}

\title{\bf Stability of discrete dark solitons \\ in nonlinear Schr\"{o}dinger lattices}

\author{D.E. Pelinovsky$^1$ and P.G. Kevrekidis$^2$ \\
{\small $^{1}$ Department of Mathematics, McMaster
University, Hamilton, Ontario, Canada, L8S 4K1} \\
{\small $^{2}$ Department of Mathematics, University of
Massachusetts, Amherst, Massachusetts, 01003-4515, USA} }

\date{\today}
\maketitle

\begin{abstract}
We obtain new results on the stability of discrete {\rm dark} solitons
bifurcating from the anti-continuum limit of the discrete nonlinear
Schr\"{o}dinger equation, following the analysis of our previous
paper [Physica D {\bf 212}, 1-19 (2005)]. We derive a criterion for
stability or instability of dark solitons from the limiting
configuration of the discrete dark soliton and confirm this
criterion numerically. We also develop detailed calculations of
the relevant eigenvalues for a number of prototypical configurations
and obtain very good agreement of asymptotic predictions with the
numerical data.
\end{abstract}

In this paper, we address the dynamical lattice model
governed by the discrete nonlinear Schr\"{o}dinger (DNLS) equation \cite{KRB}.
We consider the defocusing version of this equation in the form
\begin{equation}
\label{NLS} i \dot{u}_n + \epsilon \left( u_{n+1} - 2 u_n +
u_{n-1} \right) - |u_n|^2 u_n = 0,
\end{equation}
where $n \in \Z$, $u_n(t) : \R \to \C$, and $\epsilon > 0$. The
stationary solutions $u_n(t) = \phi_n e^{- i t}$ are found from
second-order difference equation
\begin{equation}
\label{difference} (\phi_n^2 - 1 ) \phi_n = \epsilon \left(
\phi_{n+1} - 2 \phi_n + \phi_{n-1} \right)
\end{equation}
for a real-valued sequence $\{ \phi_n \}_{n \in \mathbb{Z}}$, denoted in vector
notations by $\mbox{\boldmath $\phi$}$. Our
aim here is to study discrete {\em dark} solitons which are defined by the
non-vanishing boundary conditions at infinity, e.g. $\lim_{n \to \pm
\infty} \phi_n = \pm 1$. The limiting configuration of dark solitons
at $\epsilon = 0$ is defined by the decomposition $\mathbb{Z} = U_+
\cup U_- \cup U_0$ such that $\phi_n = \pm 1$ for $n \in U_{\pm}$
and $\phi_n = 0$ for $n \in U_0$. Our previous work \cite{PKF}
addressed stability of discrete {\em bright} solitons when ${\rm
dim}(U_+ \cup U_-) < \infty$. In this paper, we shall study
stability of discrete dark solitons, when ${\rm dim}(U_0) < \infty$
and there exists $N \geq 1$ such that $n \in U_{\pm}$ for all $\pm n
\geq N$. These solutions were considered recently in \cite{FKSF,SJ},
as well as earlier in \cite{JK,KKC} using predominantly numerical computations.

The topic of dark solitons and their stability is not only
of theoretical and mathematical interest,
but is also a subject of relevance to presently
available experimental settings. In particular, discrete dark
solitons have been observed
in the context of AlGaAs waveguide arrays in the anomalous
diffraction regime \cite{silb}. Furthermore, as was illustrated
in \cite{FKSF}, similar phenomenology can be observed in the
discrete dark solitons that arise in defocusing lithium niobate
waveguide arrays which exhibit a saturable nonlinearity due
to the photovoltaic effect; in the latter case, experimental
results were presented in the work of \cite{kip}. Although these
nonlinear optics experiments are the most promising realizations of
discrete dark solitons, such waveforms may also be relevant to
the atomic physics. In particular, dark solitons were considered
for Bose-Einstein condensates in the
presence of a periodic, so-called optical lattice, potential
\cite{konotop,markus2} (although in the latter setting, discrete
dark solitons have not yet been experimentally realized).

To determine the persistence and stability of discrete dark solitons,
we study spectra of the linear operators $L_+$ and $L_-$ defined by
\begin{eqnarray*}
(L_+ \mbox{\boldmath $\psi$} )_n & = & (3 \phi_n^2 - 1) \psi_n -
\epsilon \left(
\psi_{n+1} - 2 \psi_n + \psi_{n-1} \right), \\
(L_- \mbox{\boldmath $\psi$} )_n & = & (\phi_n^2 - 1) \psi_n -
\epsilon \left( \psi_{n+1} - 2 \psi_n + \psi_{n-1} \right).
\end{eqnarray*}
If $\mbox{\boldmath $\phi$} \in l^{\infty}(\mathbb{Z})$ for any
$\epsilon \geq 0$, then the operators $L_{\pm}$ map
$l^2(\mathbb{Z})$ to itself. Their spectrum at $\epsilon = 0$ is
computed explicitly. The operator $L_+$ has an eigenvalue $2$ of
multiplicity ${\rm dim}(U_+) + {\rm dim}(U_-) = \infty$ and an
eigenvalue $-1$ of multiplicity ${\rm dim}(U_0) < \infty$. The
operator $L_-$ has an eigenvalue $0$ of multiplicity ${\rm dim}(U_+)
+ {\rm dim}(U_-) = \infty$ and an eigenvalue $-1$ of multiplicity
${\rm dim}(U_0) < \infty$.

Since $l^2(\mathbb{Z})$ is a Banach algebra with respect to
the pointwise multiplication and the operator $L_+$ is continuously
invertible in $l^2(\mathbb{Z})$ for sufficiently small
$\epsilon \geq 0$, persistence of solutions of the
difference equation (\ref{difference}) in $l^2(\mathbb{Z}) \subset l^{\infty}(\mathbb{Z})$
with respect to small parameter $\epsilon$ is proved using the Implicit
Function Theorem. Analysis of the stability problem
\begin{equation}
\label{LL} (L_+ {\bf u})_n = - \lambda w_n, \qquad (L_- {\bf w})_n
= \lambda u_n
\end{equation}
for small $\epsilon \geq 0$ is, however, more complicated because of
the splitting of the zero eigenvalue of infinite multiplicity into a
spectral band located at
$$
\Lambda_s = \left\{ \lambda \in \mathbb{C} : \; -2 \sqrt{2\epsilon (1 + 2
\epsilon)} \leq {\rm Im} \lambda \leq 2 \sqrt{2 \epsilon (1 + 2 \epsilon)} \right\}
$$
and a number of isolated eigenvalues around the origin. We shall count these
eigenvalues by using the recent results of \cite{ChPel,PKF}.

Since $(L_- \mbox{\boldmath $\phi$})_n = 0$ and the non-decaying
sequence $\{ \phi_n \}_{n \in \mathbb{Z}}$ does not oscillate as $n
\to \pm \infty$, $0$ is at the bottom of the continuous spectrum of
$L_-$, which is located for $\lambda \in [0,4 \epsilon]$. By the
discrete Sturm theory \cite{LL}, the number of negative eigenvalues
of $L_-$ equals the number of times the sequence $\{ \phi_n \}_{n
\in \mathbb{Z}}$ changes sign on $\mathbb{Z}$. To compute this
number, we subdivide $U_0$ into segments $U_0 = \cup_{j = 1}^N
[n_j^-,n_j^+]$ for some $N < \infty$ and denote the number of
sign-changes between adjacent nodes in $U_+ \cup U_-$ by $\sigma_0$.

\begin{Lemma}
\label{lemma-count} There exists $\epsilon_0 > 0$ such that, for any
$\epsilon \in (0,\epsilon_0)$, the number of sign changes of the
sequence $\{ \phi_n \}_{n \in \mathbb{Z}}$ equals ${\rm dim}(U_0) +
\sigma_0 + \sum_{j=1}^{N} \sigma_j$, where $\sigma_j$ is associated
with the segment $U_j = [n_j^-,n_j^+] \subset U_0$, such that
\begin{equation}
\label{formula-sigma} \sigma_j = \left\{ \begin{array}{ll} 1 \;\;
\mbox{if} \;\; {\rm dim}(U_j) \;\; \mbox{is odd and} \;\; {\rm
sign}(\phi_{n_j^--1} \phi_{n_j^++1}) = 1 \\ 0 \;\; \mbox{if} \;\;
{\rm dim}(U_j) \;\; \mbox{is odd and} \;\; {\rm sign}(\phi_{n_j^--1}
\phi_{n_j^++1}) = -1 \\ 1 \;\; \mbox{if} \;\; {\rm dim}(U_j) \;\;
\mbox{is even and} \;\; {\rm sign}(\phi_{n_j^--1} \phi_{n_j^++1}) =
-1 \\ 0 \;\; \mbox{if} \;\; {\rm dim}(U_j) \;\; \mbox{is even and}
\;\; {\rm sign}(\phi_{n_j^--1} \phi_{n_j^++1}) = 1
\end{array} \right.
\end{equation}
\end{Lemma}

\begin{Proof}
By persistence of solutions in $l^{\infty}(\mathbb{Z})$-norm for
sufficiently small $\epsilon$, any sign change between the adjacent
nodes in $U_+ \cup U_-$ persists in $\epsilon$. Therefore, the
statement of the lemma is proved if we can prove
for a particular segment $U_j =
[n_j^-,n_j^+]$ of length $n_j = n_j^+ - n_j^- + 1$ that the number of
sign changes equals $n_j + \sigma_j$, where $\sigma_j$ is given by
(\ref{formula-sigma}). To do this with an application of
Lemma 2.3 from \cite{PKF}, we use the staggering
transformation $\phi_n = (-1)^n \varphi_n$ and rewrite the different
equation (\ref{difference}) in the form
$$
(1 - \varphi_n^2) \varphi_n = \epsilon \left( \varphi_{n+1} + 2
\varphi_n + \varphi_{n-1} \right).
$$
By Lemma 2.3 of \cite{PKF}, there is only one sign difference in the
sequence $\{ \varphi_n \}_{n_j^--1}^{n_j^++1}$ if ${\rm
sign}(\varphi_{n_j^--1} \varphi_{n_j^++1}) = -1$ and none if ${\rm
sign}(\varphi_{n_j^--1} \varphi_{n_j^++1}) = 1$. If $n_j$ is odd,
the staggering transformation gives $(n_j + 1)$ sign differences in
the sequence $\{ \phi_n \}_{n_j^--1}^{n_j^++1}$ if ${\rm
sign}(\phi_{n_j^--1} \phi_{n_j^++1}) = 1$ and $n_j$ sign differences
if ${\rm sign}(\phi_{n_j^--1} \phi_{n_j^++1}) = -1$. If $n_j$ is
even, however, the sequence $\{ \phi_n \}_{n_j^--1}^{n_j^++1}$ has
$n_j$ sign differences if ${\rm sign}(\phi_{n_j^--1} \phi_{n_j^++1})
= 1$ and $(n_j+1)$ sign differences if ${\rm sign}(\phi_{n_j^--1}
\phi_{n_j^++1}) = -1$. Thus, formula (\ref{formula-sigma}) is proved.
\end{Proof}

\begin{Corollary}
\label{corollary-edge-bifurcation} The number $N_0 = \sigma_0 +
\sum_{j=1}^{N} \sigma_j$ equals the number of small negative
eigenvalues of $L_-$ for $\epsilon > 0$ bifurcating from the zero
eigenvalue of infinite multiplicity for $\epsilon = 0$.
\end{Corollary}

\begin{Proof}
This follows from the fact that $L_-$ has ${\rm dim}(U_0)$ negative
eigenvalues at $\epsilon = 0$.
\end{Proof}

\begin{Definition}
\label{definition-krein-signature} The stability problem (\ref{LL}) is said
to have a purely imaginary eigenvalue of negative Krein signature if
$(L_- {\bf u}, {\bf u}) = (L_+^{-1} {\bf w},{\bf w}) \leq 0$ for the corresponding
eigenvector $({\bf u},{\bf w})$.
\end{Definition}

\begin{Theorem}
\label{proposition-stability} There exists $\epsilon_0
> 0$ such that, for any $\epsilon \in (0,\epsilon_0)$, the
stability problem (\ref{LL}) has exactly ${\rm dim}(U_0)$ pairs of
purely imaginary isolated eigenvalues of negative Krein signature
bounded away from the continuous spectrum and exactly $N_0$ pairs of
small real eigenvalues.
\end{Theorem}

\begin{Proof}
Since $L_+$ is invertible for sufficiently small $\epsilon$, we
rewrite the stability problem (\ref{LL}) in the form
\begin{equation}
\label{diagonalization} L_- {\bf w} = \gamma L_+^{-1} {\bf w}, \quad
\gamma = -\lambda^2.
\end{equation}
Since $L_-$ is not invertible for any $\epsilon \geq 0$,
we shift the eigenvalue problem to the form
$$
\left( L_- + \delta L_+^{-1} \right) {\bf w} = (\gamma + \delta)
L_+^{-1} {\bf w},
$$
for sufficiently small $\delta > 0$. Since ${\rm Null}(L_-)$ lies in
the positive subspace of $L_+^{-1}$ at $\epsilon = 0$ and the number
of negative eigenvalues of $L_-$ is unchanged in $\epsilon \in
(0,\epsilon_0)$, for a fixed $\epsilon \in (0,\epsilon_0)$, there is
a small $\delta = \delta(\epsilon)$, such that the number of
negative eigenvalues of $L_- + \delta L_+^{-1}$ is the same as that
of $L_-$. Conditions of Theorem 1 of \cite{ChPel} are now satisfied
and we count the negative eigenvalues of $L_- + \delta L_+^{-1}$
(same as for $L_-$) and $L_+^{-1}$ (same as for $L_+$) as follows:
$$
n(L_-) = N_p^- + N_n^+ + N_{c^+}, \qquad n(L_+) = N_n^- + N_n^+ +
N_{c^+},
$$
where $n(L_{\pm})$ denotes the number of negative eigenvalues of
$L_{\pm}$, $N_{c^+}$ denotes the number of complex eigenvalues
$\gamma$ in the upper half-plane, $N_n^{\pm}$ denote the number of
positive/negative eigenvalues $\gamma$ with $({\bf w}, L_+^{-1} {\bf
w}) \leq 0$ for corresponding eigenvectors ${\bf w}$, and $N_p^-$
denotes the number of negative eigenvalues $\gamma$ with $({\bf w},
L_+^{-1} {\bf w}) \geq 0$ for corresponding eigenvectors ${\bf w}$.
By Lemma \ref{lemma-count} and Corollary
\ref{corollary-edge-bifurcation}, we have $n(L_-) = {\rm dim}(U_0) +
N_0$ and $n(L_+) = {\rm dim}(U_0)$ for sufficiently small $\epsilon
\in (0,\epsilon_0)$.

At $\epsilon = 0$, there exists ${\rm dim}(U_0)$ eigenvalues $\gamma
= 1$ with $({\bf w}, L_+^{-1} {\bf w}) < 0$, where the sequence $\{
w_n \}_{n \in \mathbb{Z}}$ for the eigenvector ${\bf w}$ is
compactly supported in $U_0$. By Proposition 5.1 in \cite{ChPel},
the eigenvalue $\gamma = 1$ is hence semi-simple (that is algebraic
and geometric multiplicities coincide). Therefore, all ${\rm
dim}(U_0)$ eigenvalues persist for positive values of $\gamma$ for
sufficiently small $\epsilon$. By continuity of the eigenvectors
${\bf w}$ in $\epsilon$, the inequality $({\bf w}, L_+^{-1} {\bf w})
< 0$ holds for sufficiently small $\epsilon > 0$, and therefore,
$n(L_+) = {\rm dim}(U_0) = N_n^+$, such that $N_{c^+} = N_n^- = 0$
and $N_p^- = N_0$. Therefore, all $N_0$ bifurcations of small
negative eigenvalues of $L_-$ for $\epsilon \in (0,\epsilon_0)$ from
the zero eigenvalue of $L_-$ for $\epsilon = 0$ result in pairs of
small real eigenvalues $\lambda = \pm \sqrt{-\gamma}$ of the
stability problem (\ref{LL}).
\end{Proof}

\begin{Remark}
\label{remark-eigenvalues}
Small eigenvalues of the operator $L_-$ can be found from the difference eigenvalue problem
\begin{equation}
\label{difference-L} V_n \psi_n - \epsilon \left( \psi_{n+1} + \psi_{n-1} - 2 \psi_n \right)
= \mu \psi_n, \qquad V_n = V_n^{(0)} + \sum_{k = 1}^{\infty} \epsilon^k V_n^{(k)},
\end{equation}
where $V_n^{(0)} = (\phi_n^{(0)})^2 - 1$, $V_n^{(1)} = 2
\phi_n^{(0)} \phi_n^{(1)}$, $V_n^{(2)} = 2
\phi_n^{(0)} \phi_n^{(2)} + \left( \phi_n^{(1)} \right)^2$ and so on, due to analytic
dependence of the solution $\mbox{\boldmath $\phi$}$ of the
difference equation (\ref{difference}) on $\epsilon$.
If ${\bf w}$ is supported in $U_+ \cup U_-$ and $\epsilon = 0$,
then $L_+ {\bf w} = 2 {\bf w}$. Since $l^2$-eigenvectors of the difference equation (\ref{difference-L})
for small negative eigenvalues $\mu$ are supported in $U_+ \cup U_-$ as $\epsilon \to 0$,
a small negative eigenvalue $\mu$ for $L_-$ is related to a small
negative eigenvalue $\gamma$ for $L_+ L_-$ (that is the eigenvalue of the
stability problem (\ref{diagonalization}) with the $l^2$-eigenvector)
by the asymptotic approximation $\lim_{\epsilon \to 0} \gamma / \mu = 2$.
\end{Remark}

As the simplest application of our results, we consider two basic
configurations of discrete dark solitons from \cite{FKSF,JK,KKC}.

\begin{figure}[tbp!]
\begin{center}
\epsfxsize=7.0cm
\epsffile{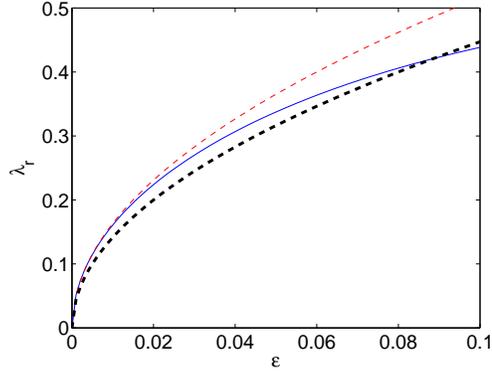}
\caption{Numerical approximations of the unstable eigenvalue for the inter-site dark
soliton (solid line) together with the asymptotic prediction
$\lambda = \sqrt{\frac{8 \epsilon}{3}}$ (dashed curve) and
the approximation $\lambda = \sqrt{2 \epsilon}$ of \cite{FKSF} (thick dashed curve).}
\label{config0}
\end{center}
\end{figure}

\begin{itemize}
\item If $U_{\pm} = \Z_{\pm}$ and $U_0 = \{0\}$ (a so-called
{\em on-site} dark soliton), then $N_0 = 0$ and, according to
Theorem \ref{proposition-stability}, the dark soliton is stable with
a single pair of purely imaginary eigenvalues of negative Krein
signature near $\lambda = \pm i$.

\item If $U_+ = \Z_+$, $U_- = \Z_- \cup \{ 0 \}$, and $U_0 =
\varnothing$ (a so-called {\em inter-site} dark soliton), then $N_0
= \sigma_0 = 1$ and the dark soliton is unstable with a single pair
of real eigenvalues. The asymptotic approximation of the unstable
eigenvalue can be obtained with the following argument. The solution
of the difference equation (\ref{difference}) is expanded in the
power series $\mbox{\boldmath $\phi$} =
\mbox{\boldmath $\phi$}^{(0)} + \epsilon \mbox{\boldmath
$\phi$}^{(1)} + {\rm O}(\epsilon^2)$, where $\mbox{\boldmath
$\phi$}^{(1)}$ is compactly supported with $\phi_0^{(1)} = 1$,
$\phi_1^{(1)} = -1$ and $\phi_n^{(1)} = 0$ for all $n \in \mathbb{Z}
\backslash \{ 0,1\}$. Since $V_n^{(0)} = 0$ for all $n \in \mathbb{Z}$
and $V_n^{(1)} = -2$ for $n = \{ 0,1\}$ and $V_n^{(1)} = 0$ otherwise,
the potential $V$
of the discrete Schr\"{o}dinger equation (\ref{difference-L}) is negative at the leading order.
By the discrete Sturm theory \cite{LL},
it traps a unique negative eigenvalue with the symmetric eigenfunction
$\psi_n = \psi_{-n+1}$, $n \in \mathbb{N}$. Using the
parametrization
\begin{equation}
\label{parameterization} \mu = \epsilon \left( 2 - e^{\kappa} -
e^{-\kappa} \right)
\end{equation}
and solving the eigenvalue problem for the eigenvector $\psi_1 = 1$,
$\psi_n = C e^{-\kappa (n-2)}$ for $n \geq 2$, we obtain $C = e^{-\kappa}$ and
$e^{\kappa} = 3$ at the leading order of ${\rm O}(\epsilon)$, which gives
$\mu = -\frac{4}{3} \epsilon + {\rm O}(\epsilon^2)$. Using Remark \ref{remark-eigenvalues},
we conclude that the pair of real eigenvalues of the
stability problem (\ref{LL}) is given by $\lambda = \pm \sqrt{-\gamma} =
\pm \sqrt{\frac{8 \epsilon}{3}} (1 + {\rm O}(\epsilon))$. This approximation is
shown on Fig. \ref{config0} with thin dashed line, while the solid line
shows results of numerical approximations of eigenvalues of the truncated
linear stability system (\ref{LL}).  It should be noted that the earlier work of \cite{FKSF}
approximated the real eigenvalue pair of the inter-site dark
soliton as $\lambda = \pm \sqrt{2 \epsilon}$. As can be readily
observed from a solid dashed line in Fig. \ref{config0}, this asymptotic
prediction is incorrect.
\end{itemize}

To illustrate more complicated applications of Theorem \ref{proposition-stability}, we
consider several composite discrete dark solitons, some of which
were studied in \cite{SJ}.

\begin{itemize}
\item If $U_+ = \{ 0 \} \cup \mathbb{Z}_+ \backslash \{ 1 \}$,
$U_0 = \varnothing$, and $U_- = \mathbb{Z}_- \cup \{ 1 \}$, then
$N_0 = \sigma_0 = 3$, such that three pairs of real (unstable)
eigenvalues occurs in the linearized problem (\ref{LL}). To find
asymptotic approximations of these eigenvalues, we again consider
eigenvalues of $L_-$ in the difference equation (\ref{difference-L})
with the potentials $V_n^{(0)} = 0$ for all $n \in \mathbb{Z}$ and
$$
\phi_n^{(1)} = \left\{ \begin{array}{ll} \phantom{tt} 1, \;\; n = -1
\\ -2, \;\; n = 0 \\ \phantom{tt} 2, \;\; n = 1 \\ -1, \;\; n = 2 \\ \phantom{tt} 0, \;\;
\mbox{otherwise}
\end{array}
\right., \qquad V_n^{(1)} = 2 \phi_n^{(0)} \phi_n^{(1)} = \left\{ \begin{array}{ll} -4, \;\; n = \{ 0,1 \} \\
-2, \;\; n = \{ -1,2\}, \\ \phantom{tt} 0, \;\; \mbox{otherwise}
\end{array} \right.
$$
We construct two symmetric eigenvectors and one anti-symmetric
eigenvector for three negative eigenvalues $\mu$. For symmetric
eigenvectors, $\psi_n = \psi_{-n+1}$, $n \in \mathbb{Z}$ with
$\psi_1 = 1$, $\psi_2 = B$, $\psi_n = C e^{-\kappa (n-3)}$, $n \geq
3$, we use the parametrization (\ref{parameterization}) and obtain
at the leading order of ${\rm O}(\epsilon)$:
$$
C = B e^{-\kappa}, \quad (e^{\kappa} - 2) B = 1, \quad B =
e^{\kappa} + e^{-\kappa} - 5.
$$
Eliminating $B$, we obtain a cubic equation for $z = e^{\kappa}$:
\begin{equation}
\label{cubic1} z^3 - 7 z^2 + 10 z - 2 = 0.
\end{equation}
There exist two solutions of the cubic equation in the interval $z >
1$, namely $z_1 \approx 1.63667$ and $z_2 \approx 5.12489$. For the
anti-symmetric eigenvector, $\psi_n = -\psi_{-n+1}$, $n \in
\mathbb{Z}$ with $\psi_1 = 1$, $\psi_2 = B$, $\psi_n = C e^{-\kappa
(n-3)}$, $n \geq 3$, we obtain at the leading order of ${\rm
O}(\epsilon)$:
$$
C = B e^{-\kappa}, \quad (e^{\kappa} - 2) B = 1, \quad B =
e^{\kappa} + e^{-\kappa} - 3.
$$
Eliminating $B$, we obtain a cubic equation for $z = e^{\kappa}$:
\begin{equation}
\label{cubic2} z^3 - 5 z^2 + 6 z - 2 = 0.
\end{equation}
Since one root is $z = 1$, we can find a unique root of the cubic
equation in the interval $z > 1$, namely $z_3 = 2 + \sqrt{2}$. Each
of the three roots above generates a negative eigenvalue for $\mu =
\epsilon (2 - z - z^{-1}) + {\rm O}(\epsilon^2)$.
Each negative eigenvalue $\mu$ of $L_-$ generates a negative
eigenvalue $\gamma$ of $L_+ L_-$ with the correspondence $\gamma = 2
\mu + {\rm O}(\epsilon^2)$. Summarizing, the three pairs of real
eigenvalues are given asymptotically by
$\lambda = \pm 0.70380 \sqrt{\epsilon}$,
$\lambda = \pm 1.84776 \sqrt{\epsilon}$ and
$\lambda = \pm 2.57683 \sqrt{\epsilon}$. These theoretical predictions
are compared with the results of full numerical linear stability
analysis in the left panel of Fig. \ref{config1}, yielding a good agreement for small
values of $\epsilon$. A typical example of the discrete dark soliton and
its linearization spectrum for $\epsilon=0.05$ is shown in the right panel
of Fig. \ref{config1}.

\begin{figure}[tbp!]
\begin{center}
\epsfxsize=7.0cm
\epsffile{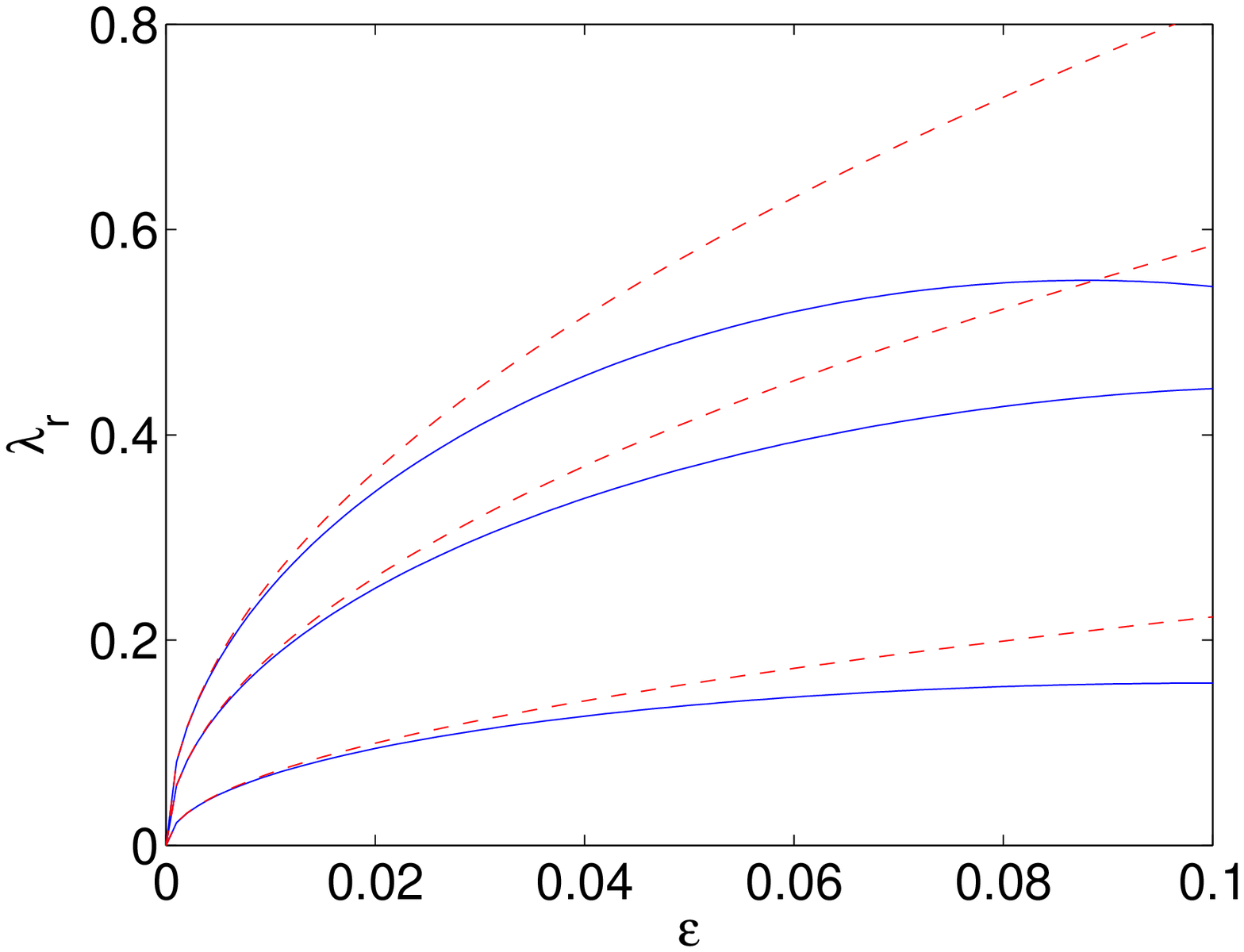}
\epsfxsize=7.0cm
\epsffile{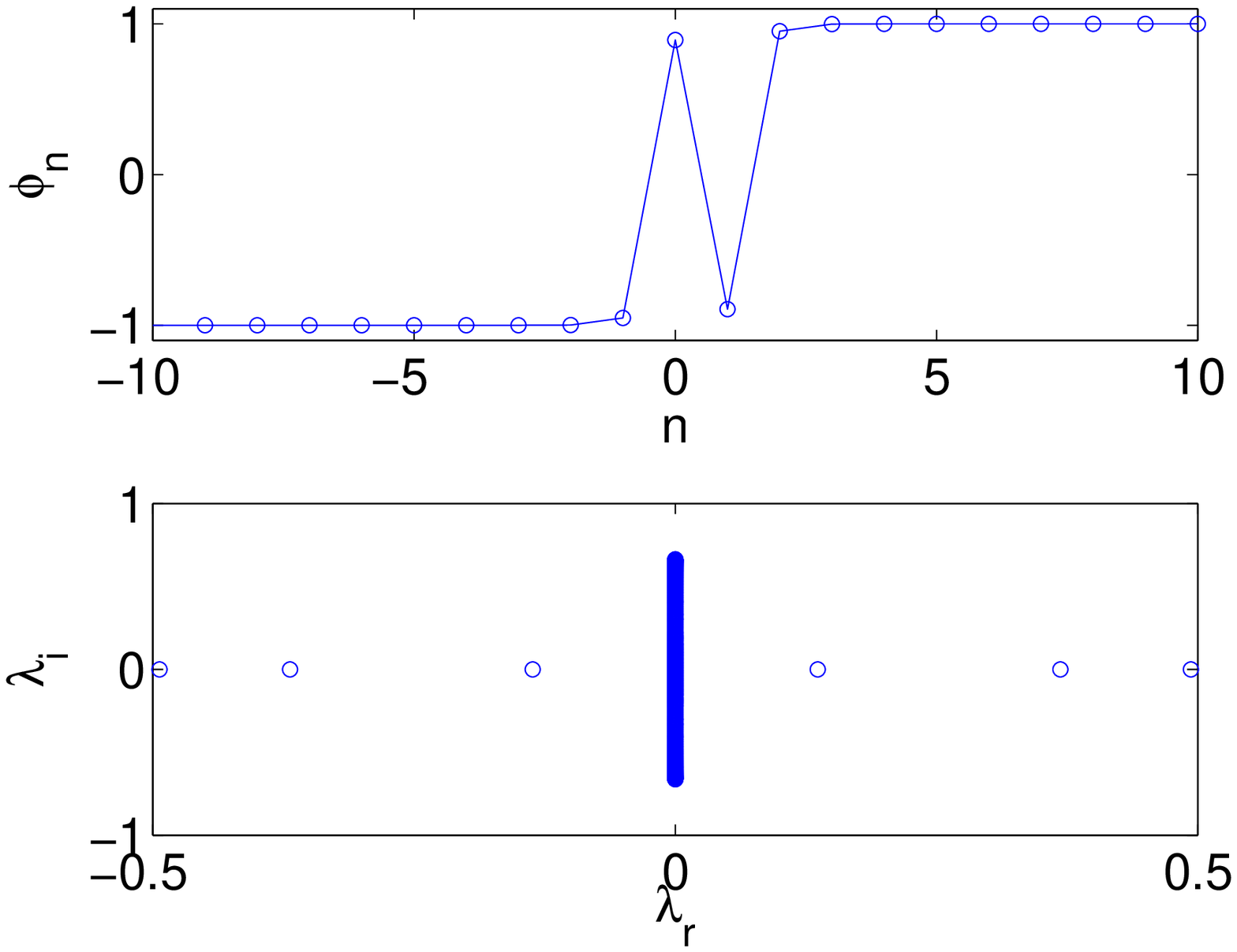}
\caption{The left panel compares the asymptotic predictions
(dashed lines) with the results of numerical linear stability analysis
(solid lines) for the three positive real eigenvalues of the
discrete dark soliton with
$U_+ = \{ 0 \} \cup \mathbb{Z}_+ \backslash \{ 1 \}$,
$U_0 = \varnothing$, and $U_- = \mathbb{Z}_- \cup \{ 1 \}$. The right
panel shows a typical example of the solution profile
(top) and the corresponding spectral plane
$\lambda=\lambda_r + i \lambda_i$ of its linearization spectrum
for $\epsilon = 0.05$.}
\label{config1}
\end{center}
\end{figure}

\item If $U_+ = \mathbb{Z}_+ \backslash \{ 1 \}$, $U_0 = \{ 0 \}$,
and $U_- = \mathbb{Z}_- \cup \{ 1 \}$, then $\sigma_0 = 1$,
$\sigma_1 = 1$, such that $N_0 = 2$ and two pairs of real (unstable)
eigenvalues occur in the linearized problem (\ref{LL}), while one
pair of imaginary eigenvalues of negative Krein signature persists
on the imaginary axis near $\lambda = \pm i$. To compute the small
negative eigenvalues of $L_-$, we compute the leading-order potential
$V_n^{(0)} = -1$ for $n = 0$ and $V_n^{(0)} = 0$ otherwise and then proceed
with the first-order potential:
$$
\phi_n^{(1)} = \left\{ \begin{array}{ll} \phantom{tt} \frac{1}{2},
\;\; n = -1 \\ \phantom{tt} 2, \;\; n = 0 \\ \phantom{tt}
\frac{3}{2}, \;\; n = 1 \\ -1, \;\; n = 2 \\ \phantom{tt} 0, \;\;
\mbox{otherwise}
\end{array}
\right., \qquad V_n^{(1)} = 2 \phi_n^{(0)} \phi_n^{(1)} = \left\{
\begin{array}{ll} -1, \;\; n = -1 \\ -3, \;\; n = 1
\\ -2, \;\; n = 2 \\ \phantom{tt} 0, \;\; \mbox{otherwise}
\end{array}  \right.
$$
Since the potential has no symmetry, we have to find the eigenvector
of the eigenvalue problem (\ref{difference-L}) in the most general
form $\psi_n = A e^{\kappa (n+2)}$, $n \leq -2$, $\psi_{-1} = B$,
$\psi_0 = C$, $\psi_1 = D$, $\psi_2 = E$, and $\psi_3 = F e^{-\kappa
(n-3)}$, $n \geq 3$. Using the parametrization
(\ref{parameterization}), we obtain at the leading order of ${\rm
O}(\epsilon)$:
$$
F = E e^{-\kappa}, \quad (e^{\kappa} - 2) E = D, \quad C + E =
\left( e^{\kappa} + e^{-\kappa} - 3 \right) D, \quad A = B
e^{-\kappa}, \quad e^{\kappa} B = B + C.
$$
Since the equation at $n = 0$ implies that $C = -\epsilon (B + D) +
{\rm O}(\epsilon^2)$, the chain of equations is uncoupled at the
variables $(D,E,F)$ and $(A,B)$ at the leading order. Working with the chain
for $(D,E,F)$, we obtain the same cubic equation
(\ref{cubic2}) with the same root $e^{\kappa} = 2 + \sqrt{2}$, which
gives the approximation $\mu = -\epsilon (1 + \frac{1}{\sqrt{2}}) +
{\rm O}(\epsilon^2)$. Working with the chain for $(A,B)$, we obtain
the equation $e^{\kappa} = 1$, which hides a small root $\kappa = {\rm O}(\epsilon)$.
To unveil this hidden eigenvalue, we have to extend the potential to
the second order by
$$
\phi_n^{(2)} = \left\{ \begin{array}{ll} \phantom{tt} \frac{1}{4},
\;\; n = \{-2,2\} \\ \phantom{tt}  \frac{7}{8}, \;\; n = -1 \\
\phantom{tt} 2, \;\; n = 0 \\ \phantom{tt}  \frac{19}{8}, \;\; n = 1
\\ -\frac{1}{2}, \;\; n = 3 \\
\phantom{tt} 0, \;\; \mbox{otherwise}
\end{array}
\right., \qquad V_n^{(2)} = 2 \phi_n^{(0)} \phi_n^{(2)} +
(\phi_n^{(1)})^2 =
\left\{ \begin{array}{ll} -\frac{1}{2}, \;\; n = -2 \\
-\frac{3}{2}, \;\; n = -1, \\ \phantom{tt}  4, \;\; n = 0
\\ -\frac{5}{2}, \;\; n = 1 \\ \phantom{tt}  \frac{3}{2}, \;\; n = 2
\\ -1, \;\; n = 3 \\ \phantom{tt} 0, \;\; \mbox{otherwise}
\end{array} \right.
$$
Using the parametrization (\ref{parameterization}) and the
representation of the eigenvector in the form $\psi_n = A e^{\kappa
(n+3)}$, $n \leq -3$, $\psi_{-2} = B$, $\psi_{-1} = C$, $\psi_0 =
D$, $\psi_1 = E$, $\psi_2 = F$, $\psi_3 = G$, and $\psi_4 = H
e^{-\kappa (n-4)}$, $n \geq 4$, we obtain at the leading order of
${\rm O}(\epsilon) + {\rm O}(\epsilon^2)$:
$$
H = G e^{-\kappa}, \;\; (e^{\kappa} - \epsilon) G = F, \;\; E + G =
\left( e^{\kappa} + e^{-\kappa} - 2 + \frac{3\epsilon}{2} \right) F,
\;\; D + F = \left( e^{\kappa} + e^{-\kappa} - 3 -
\frac{5\epsilon}{2} \right) E,
$$
and
$$
A = B e^{-\kappa}, \quad ( e^{\kappa} - \frac{\epsilon}{2} ) B = C,
\quad B + D = \left( e^{\kappa} + e^{-\kappa} - 1 -
\frac{3\epsilon}{2} \right) C,
$$
where the approximation $D = -\epsilon (C + E) + {\rm
O}(\epsilon^2)$ is sufficient for the purpose. Eliminating $B$, $D$,
$F$, and $G$, we obtain two equations at the leading order
$O(\epsilon)$:
$$
-\epsilon E = (e^{\kappa} - 1 - \epsilon) C, \quad -\epsilon C = (2
e^{\kappa} + e^{-\kappa} - 3 - \epsilon) E
$$
Using the asymptotic expansion $\kappa = \epsilon \kappa_1 + {\rm
O}(\epsilon^2)$, we reduce the problem to a quadratic equation for
$\kappa_1$ with two roots $\kappa_1 = 2$ and $\kappa_1 = 0$. The
non-zero root leads to the approximation $\mu = -4 \epsilon^3 + {\rm
O}(\epsilon^4)$. Each of the two roots above generates a negative
eigenvalue $\gamma$ of $L_+ L_-$ with the correspondence $\gamma = 2
\mu (1 + {\rm O}(\epsilon))$. Summarizing, the two pairs of
real eigenvalues are given asymptotically by $\lambda= \pm 1.84776 \sqrt{\epsilon}$
and $\lambda=\pm \sqrt{8 \epsilon^3}$, which are again found
in Fig. \ref{config2} to be in very good agreement with the full
numerical results.

\begin{figure}[tbp!]
\begin{center}
\epsfxsize=7.0cm
\epsffile{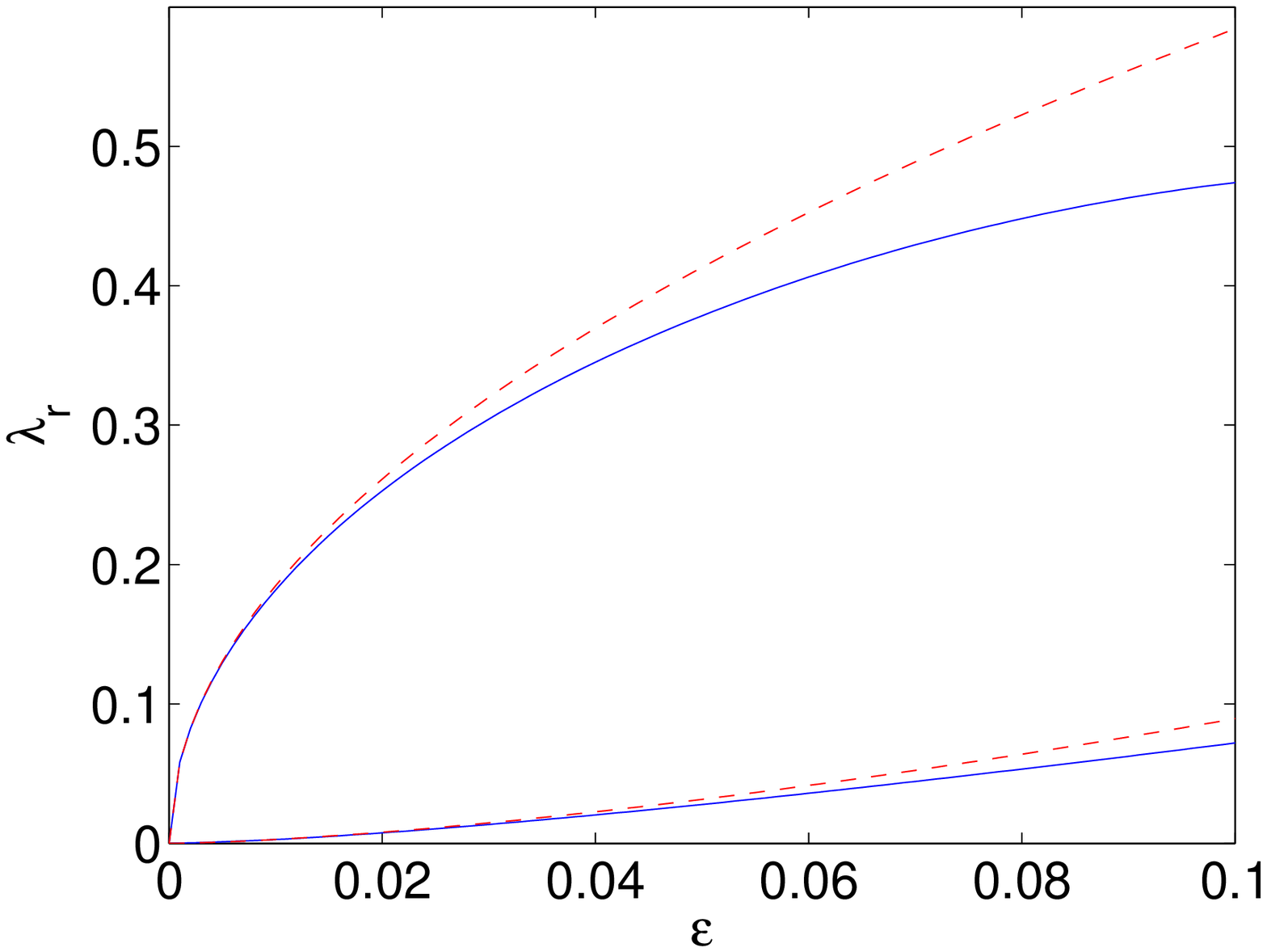}
\epsfxsize=7.0cm
\epsffile{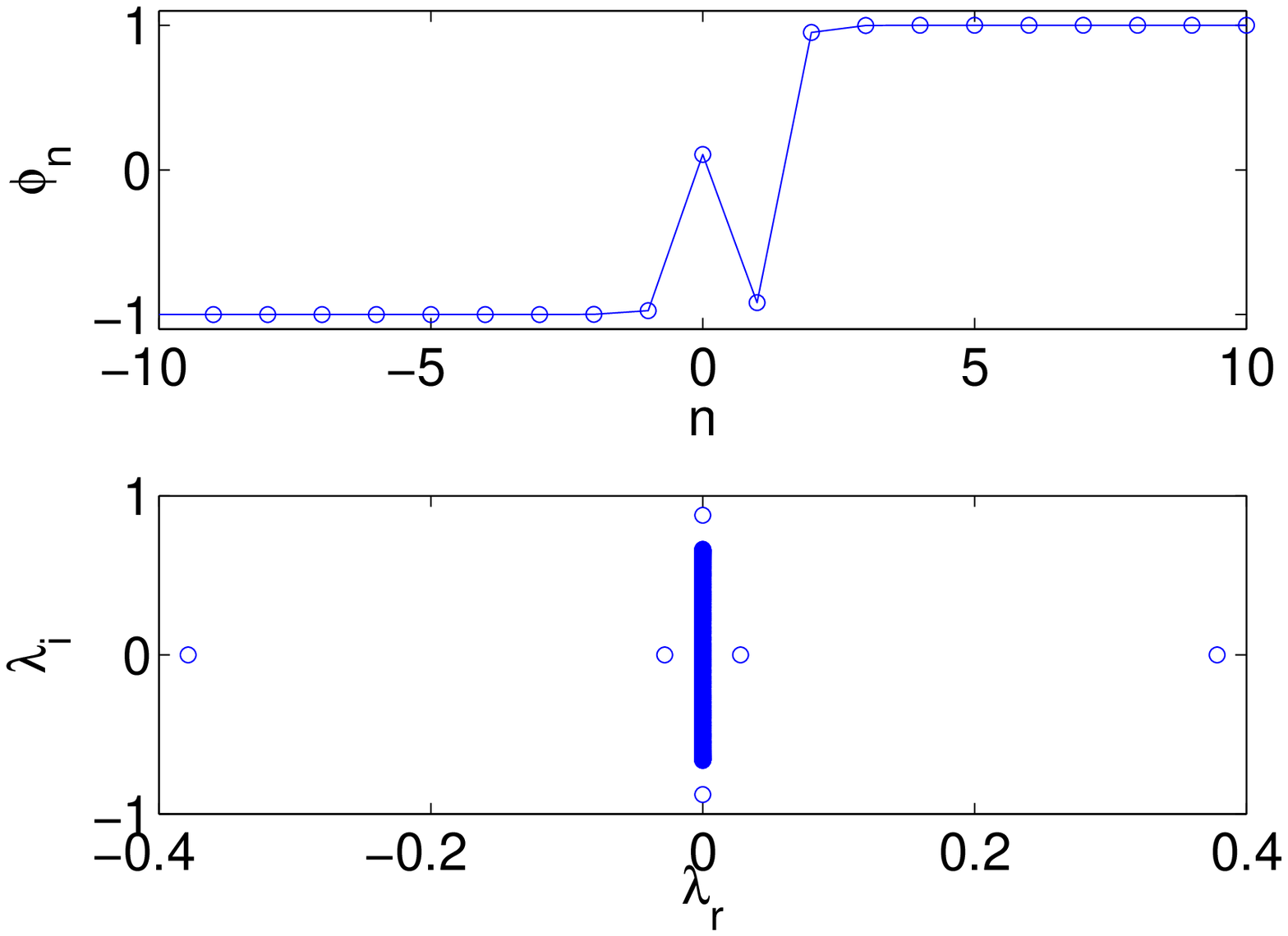}
\caption{The same as Fig. \ref{config1}, but for the discrete dark soliton
with $U_+ = \mathbb{Z}_+ \backslash \{ 1 \}$, $U_0 = \{ 0 \}$,
and $U_- = \mathbb{Z}_- \cup \{ 1 \}$.}
\label{config2}
\end{center}
\end{figure}

\item If $U_+ = \mathbb{Z}_+ \backslash \{ 1 \}$, $U_0 = \{ 0, 1
\}$, and $U_- = \mathbb{Z}_-$, then $N_0 = \sigma_1 = 1$ and one
pair of real (unstable) eigenvalues occurs in the linearized problem
(\ref{LL}), while two pairs of imaginary eigenvalues of negative
Krein signature persist on the imaginary axis near $\lambda = \pm
i$. To compute the small negative eigenvalue of $L_-$, we use Wolfram's MATHEMATICA
and compute the potentials of the eigenvalue problem (\ref{difference-L}) up to the fourth
order
$$
V_n^{(0)} = \left\{ \begin{array}{ll} -1, \;\; n = \{ 0,1 \} \\
\phantom{tt} 0, \;\; \mbox{otherwise}
\end{array} \right., \;\;
V_n^{(1)} = \left\{ \begin{array}{ll} -1, \;\; n = \{ -1,2 \} \\
\phantom{tt} 0, \;\; \mbox{otherwise}
\end{array} \right., \;\; V_n^{(2)} =
\left\{ \begin{array}{ll} -\frac{1}{2}, \;\; n = \{ -2,-1,2,3 \} \\
\phantom{tt} 1, \;\; n = \{ 0,1\}, \\
\phantom{tt} 0, \;\;
\mbox{otherwise}
\end{array} \right.
$$
and
$$
V_n^{(3)} = \left\{ \begin{array}{ll} -\frac{1}{4}, \;\; n = \{ -3,4 \} \\
\phantom{tt} \frac{1}{8}, \;\; n = \{ -2,3\}, \\
-\frac{21}{8}, \;\; n = \{ -1,2\}, \\
\phantom{tt} 5, \;\; n = \{ 0,1\}, \\
\phantom{tt} 0, \;\; \mbox{otherwise}
\end{array} \right., \qquad V_n^{(4)} =
\left\{ \begin{array}{ll} -\frac{1}{8}, \;\; n = \{ -4,5 \} \\
\phantom{tt} \frac{5}{16}, \;\; n = \{ -3,4\}, \\
- \frac{15}{8}, \;\; n = \{ -2,3\}, \\
-\frac{129}{16}, \;\; n = \{ -1,2\}, \\
\phantom{tt} \frac{45}{2}, \;\; n = \{ 0,1\}, \\
\phantom{tt} 0, \;\;
\mbox{otherwise}
\end{array} \right.
$$
Using the parametrization (\ref{parameterization}) and the symmetry
of the eigenvector $\psi_n = \psi_{-n+1}$, $n \in \mathbb{Z}$ with
$$
\psi_1 = \left\{ \begin{array}{ll} A, \;\; n = 1 \\
B, \;\; n = 2, \\
C, \;\; n = 3, \\
D, \;\; n = 4, \\
E, \;\; n = 5, \\
F e^{-\kappa (n-6)}, \;\; n \geq 6
\end{array} \right.
$$
we obtain algebraic equations for coefficients $A...F$, which are solvable
up to the the fourth order, subject to the characteristic equation
$e^{\kappa} = 1 + 2 \epsilon^2 + {\rm O}(\epsilon^3)$. Therefore, $\kappa =
2 \epsilon^2 + {\rm O}(\epsilon^3)$, such that $\mu =
-4 \epsilon^5 + {\rm O}(\epsilon^6)$. The small negative
eigenvalue of $L_+ L_-$ is thus approximated by $\gamma =
-8 \epsilon^5 + {\rm O}(\epsilon^6)$, while the pair of real
eigenvalues of the stability problem (\ref{LL}) is given by $\lambda
= \pm \sqrt{8 \epsilon^5} (1  + {\rm O}(\epsilon))$. The prediction
for this small real eigenvalue, leading to a very weak instability in
this case, is compared to numerical results in Fig. \ref{config3}.
Once again, we report very good agreement between the two.

\begin{figure}[tbp!]
\begin{center}
\epsfxsize=6.5cm
\epsffile{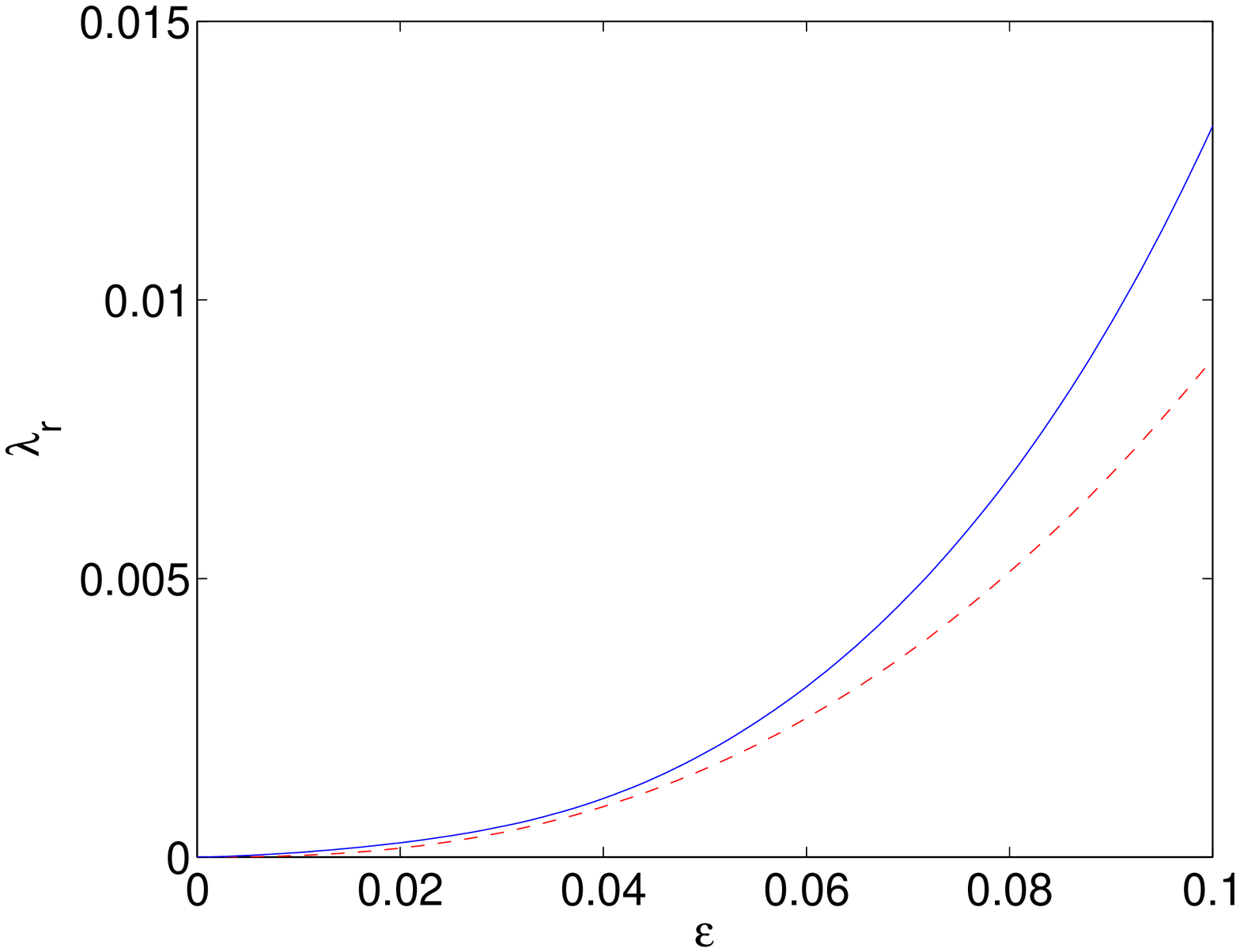}
\epsfxsize=9cm
\epsffile{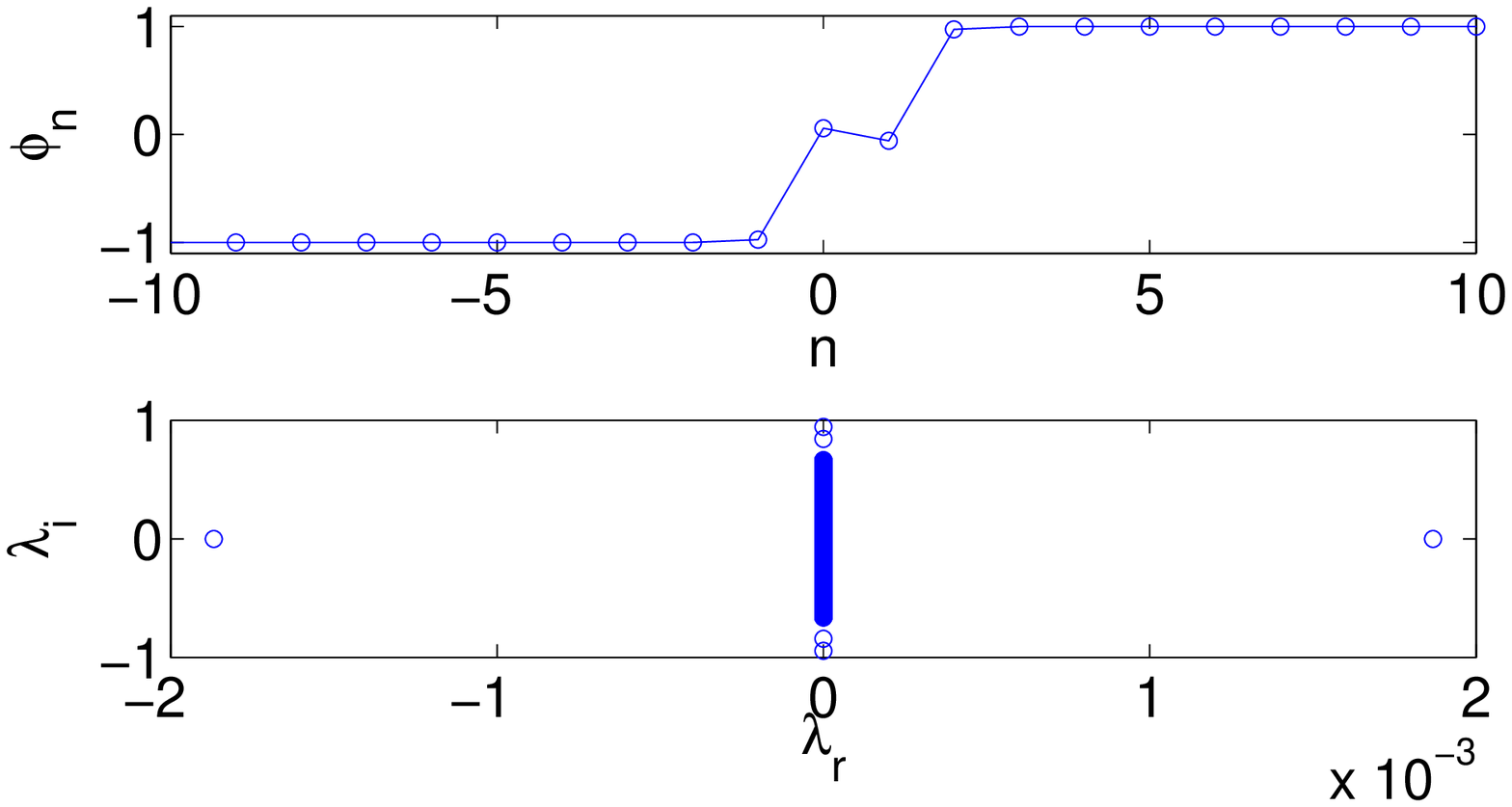}
\caption{The same as Fig. \ref{config1}, but for the discrete dark soliton
with $U_+ = \mathbb{Z}_+ \backslash \{ 1 \}$, $U_0 = \{ 0, 1
\}$, and $U_- = \mathbb{Z}_-$.}
\label{config3}
\end{center}
\end{figure}

\item If $U_+ = \mathbb{Z}_+ \backslash \{ 1, 2 \}$, $U_0 = \{ 0,
1, 2 \}$, and $U_- = \mathbb{Z}_-$, then $N_0 = 0$ and three pairs
of imaginary eigenvalues of negative Krein signature persist on the
imaginary axis near $\lambda = \pm i$. This is confirmed in
Fig. \ref{config4}, showing a typical example of the discrete dark soliton
and its linearization spectrum for $\epsilon=0.05$.

\begin{figure}[tbp!]
\begin{center}
\epsfxsize=7.0cm
\epsffile{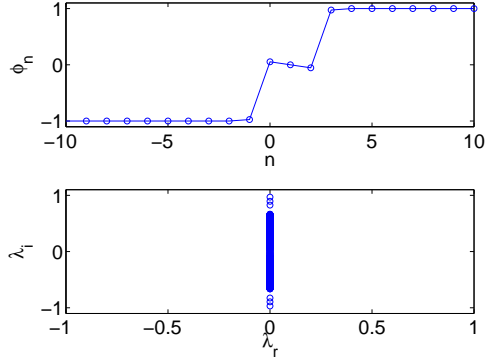}
\caption{A typical example of the solution profile (top) and the spectral
plane of its linearization spectrum (bottom) for the discrete dark soliton with
$U_+ = \mathbb{Z}_+ \backslash \{ 1, 2 \}$, $U_0 = \{ 0,
1, 2 \}$, and $U_- = \mathbb{Z}_-$ for $\epsilon=0.05$.
As predicted, the configuration is linearly stable for
small $\epsilon$, bearing three
pairs of imaginary eigenvalues (with negative Krein signature), but
no real eigenvalue pairs.}
\label{config4}
\end{center}
\end{figure}

\end{itemize}

In summary, we have offered a systematic way to assess the
stability of discrete dark solitons in the prototypical dynamical
lattice model of the DNLS equation. We have illustrated how
the number of sign changes in the limiting configuration at
the anti-continuum limit can be used to count the number
$N_0$ of small real eigenvalues of its linearization spectrum,
when deviating from the anti-continuum limit. We have also associated
the number of zeros in the limiting sequence with the number
of isolated imaginary eigenvalues of negative Krein signature
(which accounts for potential oscillatory instabilities for
larger values of the coupling).
In addition to this full characterization of the
real and imaginary eigenvalues, we have offered a systematic
approach towards computing asymptotic approximations of
the relevant eigenvalues. In particular, we
have connected small eigenvalues of operator $L_+ L_-$
to the small eigenvalues of operator
$L_-$ and have developed perturbation series expansions in
terms of the inter-site coupling constant. Within this method, relevant
computations result in a finite-dimensional matrix problem.
We have demonstrated this approach in a variety of configurations
including the on-site and inter-site dark
soliton structures of \cite{FKSF,JK,KKC}, but also in
multiple-hole configurations of \cite{SJ},
finding very good agreement between the analytical considerations
and the full numerical results. It would be of particular interest
to extend relevant computations to higher dimensional
settings, examining, for instance, the stability of
discrete defocusing vortices in the two- or three-dimensional
DNLS equations. Such considerations are deferred to future
studies.

{\bf Acknowledgements.} D.P. is supported by NSERC and PREA. 
P.K. is supported by NSF-CAREER, NSF-DMS-0505663, and NSF-DMS-0619492.


\begin{thebibliography}{99}

\bibitem{konotop} V. A. Brazhnyi and V. V. Konotop,
``Theory of nonlinear matter waves in optical lattices'',
Modern Physics Letters B  \textbf{18}, 627 (2004)

\bibitem{ChPel} M. Chugunova and D. Pelinovsky, ``Count of eigenvalues
in the generalized eigenvalue problem", arXiv:math/0602386v1 (2006)

\bibitem{FKSF} E.P. Fitrakis, P.G. Kevrekidis, H. Susanto, and D.J. Frantzeskakis,
``Dark solitons in discrete lattices: Saturable versus cubic
nonlinearities'', Physical Review E {\bf 75}, 066608-12 (2007)

\bibitem{JK} M. Johansson and Yu. S. Kivshar,
``Discreteness-Induced Oscillatory Instabilities of Dark Solitons'',
Physical Review Letters {\bf 82}, 85 (1999).

\bibitem{KRB} P.G. Kevrekidis, K.{\O}. Rasmussen and A.R. Bishop,
``The discrete nonlinear Schr{\"o}dinger Equation: a survey of
recent results'', International Journal of Modern Physics B {\bf 15}, 2833
(2001).

\bibitem{KKC} Yu. S. Kivshar, W. Kr{\'o}likowski, and O. A. Chubykalo,
``Dark solitons in discrete lattices'',
Physical Review E {\bf 50}, 5020 (1994).

\bibitem{LL} H. Levy and F. Lessman, {\em Finite Difference Equations}
(Dover, New York, 1992).

\bibitem{silb} R. Morandotti, H. S. Eisenberg, Y. Silberberg,
M. Sorel, and J. S. Aitchison,
``Self-Focusing and Defocusing in Waveguide Arrays''
Physical Review Letters {\bf 86}, 3296 (2001).

\bibitem{markus2} O. Morsch and M. Oberthaler,
``Dynamics of Bose-Einstein condensates in optical lattices'',
Reviews of Modern Physics {\bf 78}, 179 (2006).


\bibitem{PKF} D.E. Pelinovsky, P.G. Kevrekidis, and D.J. Frantzeskakis,
``Stability of discrete solitons in nonlinear Schr{\"o}dinger
lattices'', Physica D {\bf 212}, 1-19 (2005)

\bibitem{kip} E. Smirnov, C. E. R{\"u}tter, M. Stepi{\'c}, D. Kip, and V. Shandarov,
`` Formation and light guiding properties of dark solitons in one-dimensional waveguide arrays'',
Physical Review E {\bf 74}, 065601 (2006).


\bibitem{SJ} H. Susanto and M. Johansson, ``Discrete dark solitons with multiple holes'',
Physical Review E {\bf 72}, 016605 (2005)

\end{thebibliography}
\end{document}